\documentclass{elsart}
\usepackage{graphicx}

\usepackage{amssymb}
\def\gsim{\mathrel{
        \raise0.3ex\hbox{$>$}\kern-0.75em{\lower0.65ex\hbox{$\sim$}}}}

\begin{document}

\begin{frontmatter}



\title{An Efficient Numerical Scheme for Simulating Particle 
Acceleration in Evolving Cosmic-Ray Modified Shocks}


\author[label1]{T. W. Jones},
\ead{twj@astro.umn.edu}
\ead[url]{www.astro.umn.edu/$\sim$twj}
\author[label2]{Hyesung Kang}
\ead{kang@uju.es.pusan.ac.kr}

\address[label1]{University of Minnesota, Minneapolis, MN 55455, USA}
\address[label2]{Pusan National University, Pusan 609-735, Korea}


\begin{abstract}
We have developed a new, very efficient numerical scheme to solve
the CR diffusion convection equation that can be
applied to the study of the nonlinear time evolution of CR modified
shocks for arbitrary spatial diffusion properties.
The efficiency of the scheme derives from its use of 
coarse-grained finite momentum volumes. This
approach has enabled us, using $\sim 10~-~20$
momentum bins spanning nine orders of magnitude in momentum, to
carry out simulations that agree
well with results from simulations of modified
shocks carried out with our conventional finite difference
scheme requiring more than an order of magnitude more momentum points.
The coarse-grained, CGMV scheme reduces execution times by a factor 
approximately half the ratio of momentum bins used in the two methods.
Depending on the momentum dependence of the diffusion, additional
economies in required spatial and time resolution can be
utilized in the CGMV scheme, as well.
These allow a computational speed-up of at least an order of magnitude in some cases.

\end{abstract}

\end{frontmatter}

\section{Introduction}
\label{intro}
Collisionless shocks are widely thought to be effective accelerators of 
energetic, nonthermal particles (hereafter Cosmic-Rays or CRs). Those
particles play central roles in many astrophysical problems. The 
physical basis of the responsible diffusive shock acceleration (DSA) process 
is now well established through in-situ measurements of heliospheric
shocks \cite{EMP90,EBS93} and through analytic and numerical calculations 
\cite{berz94,dv81,dru83,kjg02,maldru01}.
While test particle DSA model treatments are relatively
well developed; e.g., \cite{dru83,marc99}, it has long been recognized that DSA is an integral part
of collisionless shock physics and that there are substantial and
highly nonlinear backreactions from the CRs to the bulk flows and to 
the MHD wave turbulence mediating the CR diffusive transport (see, for example, \cite{maldru01} and references therein). 
Most critically, the
CRs can capture a large fraction of the kinetic energy
dissipated across such transitions. As they diffuse upstream
the CRs form a pressure gradient that decelerates and compresses
the entering flow inside a broad shock precursor. That, in turn, can lead 
to greatly altered full shock jump conditions, especially if 
the most energetic CRs, which can have very large scattering lengths,
escape the system and carry significant energy with them. 
Also in response to the momentum dependent scattering lengths and
flow speed variations through the shock precursor the 
CR momentum distribution will take on different forms than in a simple
discontinuity.  Effective analytic (e.g., 
\cite{blas02,dv81,mal97}) and numerical (e.g., \cite{ell84}) methods have been
developed that allow one to compute steady-state modified
shock properties given an assumed diffusion behavior. On the other
hand, as the CR particle population evolves in time during the formation 
of such a shock the shock dynamics
and the CR-scattering wave turbulence evolve as well. For dynamically evolving 
phenomena, such as supernova remnants, the time scale for shock modification
can be comparable to the dynamical time scales of
the problem. 

The above factors make it essential to be able to include 
both nonlinear and time dependent effects in studies of DSA.
Generally, numerical simulations are called for.
Full plasma simulations offer the most complete time dependent treatments of
the associated shock microphysics \cite{giaetal93,scho90}, 
but are far too expensive to
follow the shock evolution over the time, length and energy
scales needed to model astrophysical CR acceleration.
The most powerful alternative approach utilizes continuum methods, with
a kinetic equation for
each CR component combined with suitably modified compressible fluid dynamical
equations for the bulk plasma (see \S 2 below).  By extending that equation set to 
include relevant wave action equations for the wave turbulence
that mediates CR transport, a self-consistent, closed
system of equations is possible (e.g., \cite{ach82,bell78,jon93,jon94})). 
Continuum DSA simulations of the kind just described are still quite 
challenging and expensive even with only one spatial dimension.
The numerical difficulty derives especially from the very large range of CR momenta that
must be followed, which usually extends to hundreds of GeV/c or 
beyond on the upper end and down to values close to those of the bulk thermal 
population, with nonrelativistic momenta. 
The latter are needed in part to account for ``injection'' of CRs due to incomplete 
thermalization that is characteristic of collisionless shocks.

One computational constraint comes from the fact that CR 
resonant scattering lengths from MHD turbulence, $\lambda(p)$, are generally
expected to be increasing functions of particle rigidity, $pc/Ze$. The
characteristic
length coupling the CRs of a given momentum, $p$, to the bulk flow and
defining the width of the modified shock precursor is the 
so-called diffusion length, $x_d(p) = \frac{1}{3}\lambda\upsilon/u_s$, where
$\upsilon$ is the CR particle speed, and $u_s$ is the bulk flow speed into
the shock. One must spatially resolve
the modified shock transition for the entire range of $x_d(p)$ in order to 
capture the physics of the shock formation and the spatial diffusion
of the CRs, in particular. The relevant 
$x_d(p)$ typically spans several orders of magnitude, beginning close to
the dissipation length of the thermal plasma, which defines the thickness of
the classical, ``viscous'' gas shock, also called the ``subshock'' in 
modified structure. This resolution requirement generally leads to
very fine spatial grids in comparison to the ``outer scale'' of the
problem, which must exceed the largest $x_d(p)$.

Two approaches have been applied successfully so far to manage this constraint
in DSA simulations. Berezhko and collaborators \cite{berz94} developed a method that
normalizes the spatial variable by $x_d(p)$ at each momentum
value of interest during solution of the CR kinetic equation. 
This approach establishes an spatial grid
that varies economically in tune with $x_d(p)$. Derived CR distribution properties at
different momenta can be combined to estimate feed-back on the bulk
flow at appropriate scales. The method was designed for use with
CR diffusion properties known {\it a priori}. 
It is not readily applied to CR diffusion behaviors incorporating 
arbitrary, nonlinear feedback between the waves and the CRs.
As an alternative that can accommodate those latter 
diffusion properties, Kang \etal \cite{kjls01} have implemented 
diffusive CR transport into a multi-level adaptive mesh refinement (AMR) environment.
The benefit of AMR in this context comes from the feature that the 
highest resolutions are only necessary very close to the subshock, which
can still be treated as a discontinuity satisfying standard Rankine-Hugoniot
relations. By efficient use of spatial gridding both of these 
computational strategies can greatly reduce the cost 
of time dependent DSA simulations.

On the other hand, the above methods do not directly address the principal
computational cost in such simulations, so they remain much more costly
compared to purely hydrodynamic or MHD simulations.
This is because the dependence of $f$ on CR momentum, $p$, adds
a physical dimension to the problem.
In practice, the spatial evolution of the kinetic equation for each CR 
constituent must be updated over the entire 
spatial grid at multiple momentum values; say, $N_p$. The value of
$N_p$ is usually large, since the spanned range of CR momentum is typically
several orders of magnitude. Physically, CRs propagate
in momentum space during DSA in response to adiabatic compression in the bulk flow,
sometimes by momentum diffusion (see, for example, equation \ref{DCE} below), or
because of various irreversible energy loss mechanisms, such as
Coulomb or radiative losses. The associated evolution rates for 
$f$ depend on the process, but generally depend on $\partial f/\partial \ln{p}$.
The conventional approach to evolving $f$
approximates $\partial f/\partial \ln{p}$ through low order finite differences
in $\Delta \ln{p}$ (e.g., \cite{berz94,fg87}).
Experience has shown that converged solutions
of $f(p)$ using such methods require $\Delta \ln{p} < 0.1$ \cite{berz94,kj91}.
In that case, for example, a mere five decades of momentum coverage
requires more than 100 grid points in $\ln{p}$.
Since spatial update of the kinetic equation at each momentum grid
point requires computational effort comparable to that for any of the accompanying 
hydrodynamical equations (e.g., the mass continuity equation), 
CR transport then dominates the computational
effort by a very large factor, commonly exceeding an order of magnitude.

An attractive alternative approach to evolving the kinetic equation 
replaces $f$ by its integral moments over a discrete set of finite 
momentum volumes, in which case $\partial f/\partial \ln{p}$ is replaced
by $f$ evaluated at the boundaries of those volumes. The method we outline
here follows that strategy. Because $f(\ln{p})$ is relatively smooth, simple subvolume
models can effectively be applied over moderately large momentum volumes.
We have found this method to give accurate solutions to
the evolution of $f$ with an order of magnitude fewer momentum
bins than needed in our previous finite difference calculations.
The computational effort to evolve the CR population
is thereby reduced to a level comparable to that for the hydrodynamics.
In recognition of its distinctive features we refer to the method as 
``Coarse-Grained Momentum finite Volume'' or ``CGMV''.
It extends related ideas
introduced in \cite{jre99}, \cite{jj99} and \cite{min01}
for test particle CR transport. 
Those previous presentations, while satisfactorily following
CR transport in many large-scale, smooth flows, did not include spatial or
momentum diffusion, so
could not explicitly follow evolution of $f$ during DSA.
Instead, analytic, test-particle solutions
for $f(p)$ were applied at shock jumps. Here we extend the CGMV
method so that it can be applied to the treatment of fully nonlinear CR modified shocks.

We outline the basic CGMV method and its implementation in Eulerian
hydrodynamics codes in \S 2. Several tests are discussed in \S 3, and
our conclusions are presented in \S 4.

\section{The Method}
\label{method}

\subsection {Basic Equation}
The standard diffusion-convection form of the kinetic equation describing the 
evolution of the isotropic
CR distribution function, $f(x,p,t)$, can be written in one spatial 
dimension as (e.g., \cite{skill75}).
\begin{equation}
\frac{\partial f}{\partial t} + u \frac{\partial f}{\partial x}
= \frac{1}{3} \left(\frac{\partial u}{\partial x}\right) \frac{\partial
f}{\partial y} +
\frac{\partial}{\partial x}\left( \kappa \frac{\partial f}{\partial x}\right )
+ \frac{1}{p^3} \frac{\partial}{\partial y}\left(p D \frac{\partial f}{\partial y}\right) +
S,
\label{DCE}
\end{equation}
where $u$ is the bulk flow speed, $y = \ln{p}$,
$\kappa$ is the spatial diffusion
coefficient, $D$ is the momentum diffusion coefficient, and
$S$ is a representative source term.
We henceforth express particle momentum in units
of $mc$, where $m$ is the particle mass.
The first RHS term in equation (\ref{DCE})
represents ``momentum advection'' in response to adiabatic compression or expansion.
For simplicity of presentation equation \ref{DCE} neglects for now
propagation of the scattering turbulence with respect to the bulk 
plasma, which can be
a significant influence when the sonic and Alfv\'enic Mach numbers of the 
flow are comparable. Although it is numerically straightforward to include this
effect, the details are somewhat complex, so we defer that to a follow-up
work focussed on CR transport in MHD shocks.

Full solution of the problem at hand requires simultaneous evolution of the
hydrodynamical flow, as well as the diffusion coefficients,
$\kappa$ and $D$. Again postponing full MHD,
the added equations to be solved are
the standard gasdynamic equations with CR pressure included.  
Expressed in conservative, Eulerian
formulation for one dimensional plane-parallel geometry, they are
\begin{equation}
{\partial \rho \over \partial t}  +  {\partial (u \rho) \over \partial x} = 0,
\label{masscon}
\end{equation}

\begin{equation}
{\partial (\rho u) \over \partial t}  +  {\partial (\rho u^2 + P_g + P_c) \over \partial x} = 0,
\label{momcon}
\end{equation}
 
\begin{equation}
{\partial (\rho e_g) \over \partial t}  +  {\partial (\rho e_g u + P_g u + P_c u) \over \partial x} = -u {{\partial P_c} \over {\partial x}}- L(x,t), 
\label{encon}
\end{equation}
where $P_{\rm g}$ and $P_{\rm c}$ are the isotropic gas and the CR pressure,
respectively, $e_{\rm g} = {P_{\rm g}}/{\rho}(\gamma_{\rm g}-1)+ u^2/2$
is the total energy density of the gas per unit mass and the rest of the 
variables have their usual meanings.
The injection energy loss term, $L(x,t)$, accounts for the
energy of the suprathermal particles transferred at low energy to the CRs.
As usual, CR inertia is neglected in such computations, since
the mass fraction of the CRs is generally tiny.
We note for completeness that $P_c$ can be computed from $f$ using the expression
\begin{equation}
P_c = \frac{4 \pi}{3} mc^2 \int^{p_{max}}_{p_{min}} p^4 f \frac{dp}{\sqrt{1 + p^2}}.
\label{PCE}
\end{equation}
In the simulations described below we set the particle mass, $m = 1$, for convenience .

\subsection {The CGVM Method}
As mentioned in \S \ref{intro}, the momentum advection and diffusion terms in equation 
\ref{DCE} typically require $\Delta y < 0.1$ when using low order 
finite difference methods in the momentum coordinate \cite{berz94,kj91}.
The resulting large number of grid points in $y$ makes finding the solution of 
equation \ref{DCE} the dominate effort in simulations of DSA.
On the other hand, previous studies of DSA as well as direct observations of 
CRs in different environments have shown that $f(p)$ is commonly well
described by the form $f \propto p^{-q(p)}$, where $q(p) = - \frac{\partial\ln{f}}{\partial y}$,
is a slowly varying function of $y$. Thus, 
we may expect a piecewise powerlaw form to provide
an efficient and accurate, two-parameter subgrid model for $f(y)$.  
Two moments of $f(p)$ are sufficient to recover the subgrid model parameters. 
We find it convenient to use
\begin{equation}
n_i = \int^{p_{i+1}}_{p_i} p^2 f(p) dp 
= \int^{y_{i+1}}_{y_i} \exp{(3y)}~f(y)d y,
\label{ni}
\end{equation}
and
\begin{equation}
g_i = \int^{p_{i+1}}_{p_i} p^3 f(p) dp
= \int^{y_{i+1}}_{y_i} \exp{(4y)}~f(y)d y.
\label{gi}
\end{equation}
The first of these moments, $n_i$,
is proportional to the spatial number density of CRs in the momentum bin 
$\Delta y_i = [y_i,y_{i+1}]$, while
for relativistic
CRs, $g_i$ is proportional to the energy density or pressure contribution
of CRs in the bin.
Then, for example,
\begin{equation}
n_i = \frac{f_i p^3_i}{q_i - 3}\left[1 - d_i^{3-q_i}\right],
\label{niq}
\end{equation}
where $f_i = f(p_i) = (p_{i+1}/p_i)^{q_i} f_{i+1}$, and $d_i = p_{i+1}/p_i$
with obvious extension to $g_i$. 
Either of these moments, plus their ratio, $g_i/p_i n_i$, can be used in
straightforward fashion (e.g., iteration) to find both
$f_i$ and the intrabin index, $q_i$.

To evolve $n_i$ and $g_i$ we
need the associated moments of equation \ref{DCE} over the finite momentum volume
bounded by $ \Delta y_i$. The result for $n_i$ is
\begin{equation}
\frac{\partial n_i}{\partial t} + u\frac{\partial n_i}{\partial x}
= F_{n_i} - F_{n_{i+1}} - n_i \frac{\partial u}{\partial x}
+ \frac{\partial }{\partial x} \left(K_{n_i} \frac{\partial n_i}{\partial
x}\right)
+ S_{n_i},
\label{DCE-n}
\end{equation}
where
$F_{n_i} = \{\dot p_i + q(p_i) D(p_i)/p_i)\} p^2_i f(p_i)$ is a flux 
in momentum space,
with
$\dot p = - p \frac{1}{3}\frac{\partial u}{\partial x}$,
and where $K_{ni}$ and $S_{ni}$ are averaged
over the momentum interval, according to
\begin{equation}
K_{n_i} = \frac{\int^{p_{i+1}}_{p_i} p^2 \kappa \frac{\partial f}{\partial x}
 dp}{\int^{p_{i+1}}_{p_i} p^2 \frac{\partial f}{\partial x} dp}
\Rightarrow \frac{\int^{p_{i+1}}_{p_i} \kappa f p^2 dp}{n_i},
\label{kni}
\end{equation}
and
\begin{equation}
S_{n_i} = \int^{p_{i+1}}_{p_i} p^2 S(p) dp.
\label{sni}
\end{equation}
Extension of the momentum flux term $F_{n_i}$ to include other processes
such as
radiative or Coulomb losses is obvious (e.g., \cite{jre99}).
In practice these fluxes should be upwinded
according to the signs of $\dot p$ and $q$. Evaluation of
$F_{n_i}$ and $q(p_i)$ at the boundaries of the included momentum
range requires application of suitable boundary conditions, of
course. We usually have set $n_{N_p+1} = 0$. In most cases
we pick a sufficiently large maximum momentum, $p_{N_p+1}$, that
this condition is important only late in the simulation, if at all.
Appropriate conditions
at the lowest momenta can be more involved,
depending on how one intends to connect the CR particle distribution
to the thermal particle distribution, as in the 
injection models discussed below.

The $g_i$-associated moment of equation \ref{DCE} is
\begin{equation}
\frac{\partial g_i}{\partial t} + u\frac{\partial g_i}{\partial x}
=F_{g_i}-F_{g_{i+1}}
+g_i\left(4\frac{\dot p_i}{p_i}+q_i\left<\frac{D}{p^2}\right>_i\right )
+ \frac{\partial }{\partial x} \left ( K_{g_i} \frac{\partial g_i}{\partial x}\right )
+ S_{g_i},
\label{DCE-g}
\end{equation}
where $F_{g_i} = p_i F_{n_i}$,
\begin{equation}
K_{g_i} = \frac{\int^{p_{i+1}}_{p_i} \kappa \frac{\partial f}{\partial x}
p^3 dp}{\int^{p_{i+1}}_{p_i} \frac{\partial f}{\partial x} p^3 dp}
\Rightarrow \frac{\int^{p_{i+1}}_{p_i} \kappa f p^3 dp}{g_i},
\label{kgi}
\end{equation}
$g_i \left<D/p^2\right>_i = \int^{p_{i+1}}_{p_i} p  D f dp$, and $S_{g_i}$
is given by an analogous expression to equation \ref{sni}.

We note that momentum binning in the CGMV scheme is quite flexible, so
that it can be easily adapted to either
uniform or nonuniform momentum bin sizes, or to a momentum range that
evolves during the simulation. We have successfully implemented both
nonuniform and evolving momentum bin structures, although, for
brevity we do not
illustrate them here.

To update the distribution function we simultaneously integrate equations (\ref{DCE-n})
and (\ref{DCE-g}) over the timestep $\Delta t^k$. Our implementation
applies these methods in Eulerian hydrodynamical codes, so we 
split the update into two parts. Spatial advection is carried out
by a second order, explicit van Leer scheme. The remaining
terms are evolved
using a second-order, semi-implicit Crank Nicholson scheme. For example,
given values $n^k_i$ at timestep $t^k$, and assuming for illustration
a uniform spatial grid, the implicit part of the solution is
given by the tridiagonal system
\begin{equation}
A_i^+n^{k+1}_{i+1} + A_i^0 n^{k+1}_i + A_i^- n^{k+1}_{i-1} = C_i^0,
\label{CNE}
\end{equation}
where
\begin{eqnarray*}
\lefteqn{A_i^+ = -\delta K_{i+1/2},}\\
\lefteqn{A_i^0 = 1 + \delta (K_{i+1/2} + K_{i-1/2}),}\\
\lefteqn{A_i^- = - \delta K_{i-1/2},}\\
\lefteqn{C_i^0 = n^k_i \left[1 - \delta (K_{i+1/2} + K_{i-1/2})\right] + n^k_{i+1} \delta K_{i+1/2} + n^k_{i-1} \delta K_{i-1/2}}\\
\lefteqn{~~~~~~~~ +\Delta t^k \left[ \bar F^k_{n_i} - \bar F^k_{n_{i-1}} + 3 n^k_i (\frac{\dot p_i}{p_i})
+ S_{n_i}\right],}\\
\label{CNC}
\end{eqnarray*}
with $\delta = (1/2) \Delta t^k/(\Delta x)^2$, and 
$K_{i+1/2} = (1/2) (K_{i+1}+K_i)$. 

The coefficients in equation \ref{CNE} are obtained with the
aid of the solutions to equations \ref{masscon} - \ref{encon},
which are updated prior to solution of equations \ref{DCE-n} and \ref{DCE-g}.
We note again that similar methods can be applied to follow the evolution of
the wave turbulence that resonantly scatters CRs and that
defines the spatial and momentum diffusion coefficients. In that case
one begins with the wave action equation for the appropriate
waves rather than the particle kinetic equation (e.g., \cite{bell78}).

The solution of equation (\ref{CNE}) for either $n_i$ or $g_i$
is quite analogous to our previous FD methods. Thus, since the CGMV
method evolves two quantities rather than one, the
relative effort required for a given $N_p$ is roughly twice in the CGMV
scheme that required in the FD scheme.  Our tests confirm this expected
scaling. On the other hand, since $N_p$ can be
dramatically reduced in a CGMV simulation the method can still be 
more efficient by a large factor.

\subsection {A Flux Fraction Injection model}

The most common source term represented by $S$ in equation (\ref{DCE}) 
is injection at the shock of low energy CRs from the thermal plasma. 
There is presently no generally accepted theory for that process. 
However, we have implemented two commonly used models successfully 
into the CGMV scheme. For completeness we outline those here.

The simplest and one of the most
frequently applied injection models assumes that a small, fixed fraction 
of the thermal particle flux through the gas subshock, $\epsilon_{inj}$,
is injected at a momentum $p_{inj} = \alpha c_{s_2}$, where $\alpha$ is
a constant greater than unity and $c_{s_2}$ is the plasma sound speed
immediately downstream of the subshock (e.g., \cite{kj91}). This gives
$S(p) = \epsilon_{inj} (\rho_1/\mu m_p)u_s w(x - x_s)\delta(p - p_{inj})$,
where $\rho_1$ is the plasma mass density just upstream of the subshock,
$\mu m_p$ is the plasma mean particle mass, $u_s$ is the subshock speed
with respect to the plasma immediately upstream and $w$ is a normalized weight
function that allows the injection to be distributed across the numerical
shock structure. In this case $S_{n_i} = \frac{1}{4\pi}\epsilon_{inj} 
(\rho_1/\mu m_p)u_s w(x - x_s)$, while $S_{g_i} = p_{inj} S_{n_i}$
in the momentum bin with $p_i < p_{inj} < p_{i+1}$. Both $S_{n_i}$  and $S_{g_i}$
are zero, otherwise.  The energy extracted from the thermal plasma is
simply $L = \frac{1}{2} \epsilon_{inj}w(x) \alpha^2 c^2_{s_2} \rho_1 u_s$.
For convenience we call this injection model the ``flux fraction'' or ``FF''
model.

\subsection {A Thermal Leakage Injection model}
A more sophisticated approach to injection physics
includes models of the
physical processes moderating particle orbits in the post shock
flow region in order to estimate the probability that particles
of a given speed will be able to escape back upstream, across the subshock.
In such ``thermal leakage'' (TL) models for CR injection at shocks, most of
the downstream thermal
protons are locally confined by nonlinear MHD waves and only particles
well into the tail of the postshock Maxwellian distribution can
leak upstream across the subshock.
In particular, ``leaking'' particles not only must have velocities large
enough to swim against the downstream flow in order to return across 
the shock, they must also avoid being scattered during that passage by
the MHD waves that mediate the plasma subshock. 
To model TL injection we utilize a ``transparency function'',
$\tau_{\rm esc}$, expressing the probability that supra-thermal
particles at a given velocity can leak upstream from behind the subshock
(see \cite{gies00} for details). In particular we set
\begin{equation}
\tau_{esc}(\upsilon,u_2)=H\left[ \tilde{\upsilon}-(1+\epsilon_B) \right]
        {(1-\frac{1}{\tilde{\upsilon}}) \over
        (1-\frac{u_2}{\upsilon}) }
 \exp\left\{ -\left[\tilde{\upsilon}-(1+\epsilon_B)\right]^{-2}\right\},
\label{tau}
\end{equation}
where $u_2$ is the postshock flow speed in the subshock frame, $H$ is the
Heaviside step function, 
and the particle velocity is normalized to $\tilde{\upsilon} = v \epsilon_B/u_2$.
The parameter, $\epsilon_B = B_0/B_{\perp}$, measures the ratio of the 
amplitude of the postshock MHD wave
turbulence $B_{\perp}$ to the general magnetic field aligned with the
shock normal, $B_0$. Both hybrid simulations and theory suggest that 
$0.25 \lesssim \epsilon_B \lesssim 0.35$ \cite{malvol98},
so that the model is well constrained. With this $\tau_{esc}$
the shock is completely ``opaque'' to
particles with momenta less than $p_1$,
{\it i.e.,} $\tau_{esc}=0$ for $p<p_1$, 
where $p_1 = m_p u_2 (1+\epsilon_B)/\epsilon_B$. 
So $p_1(t)$ is the lowest momentum of the first momentum bin
in the TL model and changes in time with the postshock flow speed.
For $\epsilon_B \sim 0.3$,
$p_1 \sim 4 - 5 m_p u_2$. The shock becomes virtually transparent
to particles with momenta two to three times greater than $p_1$. 
For strong, unmodified shocks $p_1$ in the TL model and $p_{inj}$ in
the FF model are similar when $\epsilon_B \sim 0.3$ and $\alpha \sim 2$.
Under those circumstances the initial injection rates will be roughly
similar, although differences in model physics lead to different behaviors
as such shocks become modified (see, e.g., \cite{berz95}, \cite{gies00}, \cite{kjg02}).

The TL model is implemented in the CGMV scheme by the following numerical 
approach. After solution of equations (\ref{DCE-n}) and (\ref{DCE-g})
the net changes in $n_i$ and $g_i$ are corrected (reduced) in the upstream
region by application of the transparency function as follows:  
\begin{equation}
n_i^{k+1} = n_i^{k} + \int_{p_i}^{p_{i+1}} \tau_{esc}(p) 
(\tilde f_i^{k+1} - f_i^k) p^2 dp
\label{leakn}
\end{equation}
and
\begin{equation}
g_i^{k+1} = g_i^{k} + \int_{p_i}^{p_{i+1}} \tau_{esc}(p) 
(\tilde f_i^{k+1} - f_i^k) p^3 dp
\label{leakg}
\end{equation}
where $\tilde f_i^{k+1}$, found using equation (\ref{niq}), is the CR 
distribution updated with equations (\ref{DCE-n}) and (\ref{DCE-g}).
The energy loss rate of the bulk plasma to injection into
the $i$-th CR momentum bin can be approximated by 
\begin{equation}
L_i(x,t) \approx -{ {4\pi m_p c^2} \over 3} ({{\partial u} \over{\partial x}})
\int_{p_i}^{p_{i+1}} {{\partial \tau_{esc}} \over \partial p } 
p^3(\sqrt{p^2+1}-1) f_i^ndp 
\label{leakL}
\end{equation}
(see \cite{kjg02}).

With the piece-wise power-law subgrid model ($f_i(p) = f_i (p/p_i)^{-q_i}$)
the integrals in equations (\ref{leakn})-(\ref{leakL}) can be written:
\begin{equation}
\int_{p_i}^{p_{i+1}} \tau_{esc}(p) f_i(p) p^2 dp=
{{n_i (q_i-3)} p_i^{q_i-3} \over (1-d_i^{3-q_i})} 
\int_{p_i}^{p_{i+1}} \tau_{esc} p^{(3-q_i)} dy,
\label{leakn2}
\end{equation}
\begin{equation}
\int_{p_i}^{p_{i+1}} \tau_{esc}(p) f_i(p) p^3 dp=
{{g_i (q_i-4)} p_i^{q_i-4} \over (1-d_i^{4-q_i})} 
\int_{p_i}^{p_{i+1}} \tau_{esc} p^{(4-q_i)} dy,
\label{leakg2}
\end{equation}
and
\begin{equation}
L_i(x,t) \approx -{ {4\pi m_p c^2} \over 3} ({{\partial u} \over{\partial x}})
{ {g_i p_i^{q_i-4} (q_i-4)}\over (1-d_i^{4-q_i}) }
\int_{p_i}^{p_{i+1}} {{\partial \tau_{esc}} \over \partial p }  
p^{4-q_i} (\sqrt{p^2+1}-1) dy \\
\label{leakL2}
\end{equation}
In practice, this leakage step is significant only for the lowest few momentum
bins, so that this correction need not be applied to all bins. 

\section{Discussion}
\label{disc}

In order to test the performance of the new CGMV scheme we have
installed it into two distinct one-dimensional Eulerian hydrodynamic (HD) codes 
that we have 
previously applied to studies of CR-modified shocks using conventional
finite 
difference (FD) methods to solve the diffusion convection equation.
In this section we briefly discuss the results and compare the CGMV and FD 
behaviors.
Both of the host HD codes are constructed from high order, conservative Riemann 
solver-based schemes designed to capture shocks sharply. 
First we describe results from
the CGMV scheme installed in a second order ``Total Variation Diminished'' (TVD)
HD code based on the finite difference scheme of Harten \cite{hart83}. This is the HD version
of the MHD-CR code used by us in a previous study of CR modified shocks
\cite{kj97}. 
Gas subshocks in the TVD scheme typically spread over 2-3
numerical zones. 
An outline of the code mechanics and the FD CR scheme
can be found there, in \cite{fjr95,kj91} and references cited
in those papers. The FD solver employed a
Crank-Nicholson routine originally introduced in \cite{fg87} 
for evolving $p^4 f(p)$ that is similar to equation (\ref{CNE}).
For the TVD tests we applied the FF injection model.

In addition we present CGMV tests carried out with our 
Cosmic Ray Adaptive refinement SHock (CRASH) code. CRASH
is based on the high order Godunov-like shock 
tracking algorithm of LeVeque \cite{levshy95}. The hydrodynamics
routine in that code employs a nonlinear
Riemann solver to follow shock discontinuities within the zones of an 
initially uniform grid. Thus, gas subshocks in CR-modified shocks 
remain discontinuous throughout a simulation, allowing CR transport
to be modeled down almost to the scale of the physical shock thickness. 
CRASH also employs adaptive mesh refinement (AMR) around shocks in
order to reduce the computational effort on the spatial grid. 
Refinement is centered on the subshock and each level spans 100 zones
with a resolution twice as fine as the level above it. The number of
refinement levels depends on what is required to capture diffusion of
the lowest energy CRs.
The standard, previously documented version of CRASH uses the same FD methods as the TVD code
to solve the diffusion convection equation for CR transport. It is
described in \cite{kjls01} and \cite{kjg02}. For the CRASH tests 
discussed in this paper we employed the TL injection model.

\subsection{TVD-CR Tests}

We first examine some results obtained using the TVD-CR code 
with both FD and CGMV schemes used to model the evolution of a
strong CR-modified shock.
Fig. 1 and Fig. 2 illustrate the evolution of shocks formed
by the reflection of a Mach 30 flow (adiabatic index,
$\gamma = 5/3$) off the left grid boundary. The resulting
piston-driven shock initially has a Mach number, $M_s \approx 40$.
The density and sound speed of plasma entering from
the right boundary were set to $\rho = 1$ and $c_s = \frac{1}{30}$, respectively,
so that the inflow speed was unity. The time unit for the
calculations is also set by these scalings. In order to relate
hydrodynamical variables to CR momenta it is necessary to
fix the unit flow speed (the inflow speed in all the simulations
discussed in this paper) with respect to the speed of light;
i.e., $c = 1/\beta$. We set $\beta = 10^{-2}$ in the TVD-CR simulations.
Time steps were fixed by a standard Courant condition,
$\Delta t = 0.8 \Delta x / \max ( u \pm c_s)$.

For the simulation illustrated in Fig. 1 evolution of the CR 
distribution is followed over the momentum 
range $[p_{min},p_{max}] = [2\times 10^{-4}, 1.6\times 10^3]$ 
($y_{max} - y_{min} \approx 16$).
The simulation represented in Fig. 2 included the momentum 
range $[p_{min},p_{max}] = [2\times 10^{-4}, 2.4\times 10^5]$ 
($y_{max} - y_{min} \approx 21$).

The CR diffusion coefficient, $\kappa$ is spatially uniform
and set to $\kappa(p) = 0.1 p^{0.51}$. 
In a quest for a reasonably generalized behavior that
required minimal computational effort, this choice was motivated by 
results from Malkov, who found
self-similar analytic steady-state solutions for strong CR-modified shocks
that apply to all powerlaw forms of $\kappa (p)$,
so long as the powerlaw index is steeper that $0.5$ \cite{mal99}. 
Thus, our $\kappa$ choice leads to fairly general shock behaviors in a way that
minimizes the width of the precursor. That width, which
determines the minimum space that must be simulated ahead of
the subshock, is set by 
$x_d(p_{cutoff})$, where $p_{cutoff}$ represents the maximum
momentum contained.
The TVD-CR simulations utilize a uniform, fixed grid, so, for
example, the Bohm diffusion form modeled in the CRASH simulations
below would lead to excessive costs for the TVD-CR FD test simulations
presented in this section.
The spatial resolution required for the calculations
is set by $x_d(p_{min})$, since
accurate solutions of the diffusion-convection equation
require good structural information in the diffusive shock precursor
upstream of the subshock. 
Previous convergence tests
have shown that $\Delta x < 0.1 x_d$ is desirable (e.g., \cite{kj91}).
For the problems illustrated in Fig. 1 and Fig. 2
these considerations led us to set $\Delta x = 3.8\times 10^{-4}$
for both the FD and CGMV simulations. By varying this
resolution, we verified that the
shock evolution is reasonably converged with respect to $\Delta x$.

The simulation followed in Fig. 1 assumes
a pre-existing CR population, $f(p) \propto p^{-4.5}$, corresponding
to an upstream CR pressure, $P_c= P_g$. No fresh
injection is included at the shock; i.e., $\epsilon_{inj} = 0$.  This test then provides a simple
and direct comparison between the CGMV and FD schemes for
solving the diffusion-convection equation, since it omits
any complications related to the injection model.

This simulation pair evolves the shocks
until $t = 30$, which is sufficient to accelerate CRs to
ultrarelativistic momenta. The spatial grid
spans the interval [0,16], which is sufficient to contain the 
leading edge of the CR precursor to the end of the simulation. 

The CR-modified shock spatial structures and the CR momentum
distributions at the subshock are shown in Fig. 1 at 
times $t = 2, 10, 20,~{\rm and}~30$. Before comparing the 
solutions obtained with the two different methods, it is useful to 
summarize briefly the physics captured during the shock evolution.
All the behaviors described here have been reported previously by multiple authors.
The figure shows the well-known property of strongly modified shocks that the
CRs extract most of the energy flux into the structure.
That leads to
a substantial drop in the postshock gas pressure, $P_g$, and a large increase
in the postshock density, $\rho$. Together those indicate a
strongly reduced postshock gas temperature. The decreased 
temperature is evident in the $p^4 f$ plot at the subshock, which is
dominated at low momenta by the Maxwellian distribution of the bulk
plasma.
As CRs diffuse upstream against the inflowing gas they compress
the flow within the precursor, preheating the gas (adiabatically in these
simulations). Initially, while the CR pressure is relatively small
compared to the incoming momentum flux, the gas subshock remains 
strong enough to produce a full four times density compression on top 
of the precompression.
However, the subshock weakens once $P_c \sim \rho u^2$, 
reducing the subshock 
compression in this case to a factor $\approx 2.6$, corresponding to a
subshock Mach number near 2.3.
That evolution explains the well-known, transitory ``density spike'' in the shock
structures seen after $t = 2$.
We note that since energy extracted from the flow by CRs becomes increasingly
spread upstream and downstream due to CR diffusion, the total
compression in such an {\it evolving} modified shock would always exceed the 
factor of seven one would predict for a strong, fully relativistic gas shock.
For this simulation no significant CR energy escapes the spatial grid
through upstream diffusion. However, at late times ($t \gsim 20$)
the partial pressure due to CRs just below $p_{max}$ is sufficient
that escape across the upper momentum boundary is significant. This
contributes to the slow decrease in $P_c$ behind the shock and
the increasing total compression through the
shock transition that is visible in Fig. 1.

In the early evolutionary stages of this flow, while shock modification is
modest, the CR momentum distribution resembles the powerlaw
form, $p^4 f(p) = const$,  predicted by test particle theory 
for a strong shock (e.g., \cite{dru83}).
With the spatial diffusion coefficient used in this simulation
the high momentum cutoff to the distribution
increases with time approximately as $p_{cutoff} \propto t^2$ (see, e.g., \cite{lag83}).
As shock modification intensifies, most of the flow compression shifts
from the subshock to the precursor. Then DSA of high momentum CRs 
occurs predominantly within the precursor rather than near
the less important subshock. Consequently the CR distribution develops
the familiar upwards-concave form resulting from the momentum
dependent CR diffusion length. CRs of higher momentum experience
a greater velocity jump within the precursor, so gain more energy each 
time they are reflected within the shock structure. That flattens
the distribution, $f(p)$ at momenta below $p_{cutoff}$. 
The result is a bump in the distribution of $p^4 f(p)$.
On the other hand, CRs with momenta only a little above the
injection range remain trapped close to the subshock. Their distribution
closely approaches the steady state, powerlaw, test particle form
appropriate to the weakened subshock. That feature extends
upwards in momentum as the bump near $p_{cutoff}$ moves
upwards. 

Looking finally to compare the two methods used to
evolve the shock evolution
displayed in Fig. 1 we see results from the FD
scheme with $\Delta y = 0.11$ and the CGMV scheme with $\Delta y = 1.0$.
The agreement is generally very good. All the dynamical
quantities, including shock jumps
and the CR momentum distributions show excellent agreement.
The curves representing the FD and CGMV distributions of
$\rho$, $P_g$ and $P_c$ and virtually indistinguishable in the
plots.
Most notably, all features formed in the FD evolution of the CR momentum 
distribution are faithfully reproduced by the much coarser CGMV
distribution.

Given the excellent comparison in this strongly modified
flow it is satisfying to note that
the execution time required for the CGMV solution was a little less than
20\% of that for the FD solution, demonstrating the significantly higher
efficiency of the former method. The 
speed-up observed in our implementations of the two methods
is roughly in accordance with what we would predict for a given reduction 
factor in the number of momentum values used, since the
CGMV method requires one to evolve two distributions $n_i$ and
$g_i$ for each momentum bin. We address convergence with respect
to momentum resolution in our discussion of a second shock.

Fig.2 shows a Mach 30 flow similar to that in Fig. 1. In this case
FF injection is included with commonly assumed values,
$\epsilon_{inj} = 10^{-3}$ and $\alpha = 2.0$ (see \S 2.3),
the upper momentum bound is increased to $p_{max} = 2.4\times 10^5$
and the spatial grid extends farther from the piston to $x_{max} = 25$. 
A negligible pre-existing CR population is
included to avoid numerical issues coming from the fact that our CGMV 
scheme requires computation of the ratio $g_i/p_in_i$ over the entire grid.

The simulations evolved the shock
until $t = 70$, which leads to $p_{cutoff} \sim 10^3$.
The spatial grid,
spanning the interval [0,25], is sufficient to contain the CR 
leading edge of the precursor almost to the end of the simulations. 
However, after $t \sim 50$ some CR energy
escapes through the right boundary, due to diffusion upstream, mimicking
the behavior of a ``free escape'' boundary  (FEB) (e.g., \cite{kj95} and
references therein). Just as for the shock simulated in Fig.1,
this energy loss amounts to a cooling process, so that the total shock
compression increases with time as the simulation ends.

Again the agreement between FD and CGMV simulations is very good. 
Very early in the evolution of the shock, when the CRs are dominated by 
freshly injected nonrelativistic particles,
the shock evolution is slightly faster in the CGMV scheme.
That influence becomes insignificant later on, so that the modified
shock structures found by the two schemes are almost identical
as is the distribution $p^4 f(p)$ at the shock.
There is a small residual effect that the position of the subshock
in the CGMV simulation lags slightly behind that of the FD simulation,
and that $P_c$ is slightly higher in the CGMV simulation near the 
piston, where the shock first formed.

Since the efficiency of the CGMV method comes from its ability to
cover the momentum range coarsely, it is important to evaluate
how broad the momentum bins can be and still faithfully model
the evolution of the shock.
Fig. 3 illustrates convergence of the CGMV
scheme with respect to momentum bin size, $\Delta y$, at $t= 70$
for the flow modeled in Fig. 2. The upper panel
plots the spatial $P_c$ distributions, while the lower panel shows
the particle momentum distributions at the gas subshock.
For reference the corresponding FD solution ($\Delta y = 0.11$)
is shown by the dotted red curves.
Solutions from the CGMV scheme are plotted for $\Delta y = 1$,
$\Delta y = 1.40$ and $\Delta y = 2.1$. 
Even the coarsest of the CGMV solutions is in basic agreement with the
other CGMV solutions and
with the FD solution. Fine details in the momentum distribution
are naturally obscured as the CGMV bin size increases. The largest 
bins with $\Delta y = 2.3$ span a decade in CR momentum 
($p_{i+1}/p_i = 10.2$), but still capture the
basic dynamical properties of the CRs correctly. 

However, the quality of the CGMV solutions deteriorates for still 
larger momentum bins in these experiments, once the simple subgrid
model for the momentum distribution becomes inadequate.
As illustrated in the lower
panel of Fig. 3 already momentum bins larger than roughly $\Delta y \sim 1.5$
cannot closely follow
sharper structures in the momentum distibution that develops
at the ends of the CR distribution. That enhances $P_c$ 
upstream of the subshock, where the flow is both cold and strongly
compressed as it approaches the subshock. When the errors become
excessive, for $\Delta y > 2.3$, in this case,
$P_c$-induced overcompression in this region 
can cause the Riemann solver in our TVD
code to perform poorly or even to fail in high Mach number flows. 
In general the largest allowed momentum bin size, $\Delta_{max}$
should depend on the strongest curvature
of the CR momentum distribution function as well as the degree of
shock modification.

\subsection{CRASH Tests}

For a third test example we illustrate in
Figs. 4 and 5 simulation results using the CRASH code with
the TL injection model applied to a Mach 10
flow reflecting off the left computational boundary.
The initial gas shock Mach number is approximately 13.
As for the previous test, the upstream gas density and 
flow speed are set to unity, with upstream sound speed, 
$c_s = \frac{1}{10}$, and $\beta = 5\times 10^{-3}$. The
time unit is defined accordingly. 
CR momenta are tracked over the range $[p_1,10^4]$,
where $p_1(t)$ is the smallest momentum that can leak upstream
(see equation 15).

In this case a Bohm-type diffusion model with 
$\kappa_B= p^2/\sqrt{p^2+1}$, is
adopted and the TL injection parameter, $\epsilon_B=0.2$ is used.
The CRASH test was significantly more computationally
demanding than the TVD-CR tests.
Note first that in the CRASH simulation the
value of $\kappa_B(p=1)=1/\sqrt{2}$, while
$\kappa(p=1)=0.1$ in the previous examples shown in Figs. 1-3.
Consequently, $x_d(p=1)$ is about seven times greater in the
current case, and the nominal physical scale of
the precursor and its formation timescale are similarly lengthened.
In addition, the stronger momentum dependence of Bohm diffusion 
coefficient means that the precursor width expands more strongly
as $p_{cutoff}$ increases.  The associated time rate of increase in
$p_{cutoff}$ is, however, slower, so that the shock must evolve
longer to reach a given $p_{cutoff}$. These factors substantially
increase the size of the physical domain needed to reach a given $p_{cutoff}$.

Fig. 4 shows the early evolution of this CR-modified 
shock for $t \leq 20$ as computed with both the FD and the CGMV methods.
The spatial domain for this simulation is [0,20]. 
The base spatial grid included
$10^4$ zones, giving $\Delta x_0 = 2\times10 ^{-3}$.
Since it is necessary to resolve
structures near the subshock on scales of the diffusion length for
freshly injected, suprathermal CRs, the AMR feature of the
CRASH code is utilized. 
The FD simulation is carried out with 7 refined grid levels;
four levels of refined grid are applied in the CGMV simulation.
240 momentum points ($\Delta y = 0.058$)are used in the FD simulation, 
while the CGMV simulation includes 20 momentum bins ($\Delta y = 0.72$).
The time step for each refinement level, $\Delta t_{l_g}$, is determined by a standard
Courant condition, that is, $\Delta t_{l_g} = 0.3 \Delta x_{l_g} / \max
( u \pm c_s)$.
Although the Crank-Nicholson scheme is stable with an arbitrary time
step, the diffusion convection equation is solved with the time step 
smaller than $\Delta t_{DC,l_g} \sim 2 \Delta y (\Delta x_{l_g}/ u_s$) 
to maintain good accuracy in the momentum space advection
({\it i.e.,} $dy/dt = - \frac{1}{3}\frac{\partial u}{\partial x}$). 
With $\Delta y = 0.058$, the required time step is smaller by a factor of three or so
than the hydrodynamic time step in the FD simulation. 
Consequently, the FD diffusion convection solver is typically subcycled about 3 times 
with $\Delta t_{DC,l_g}$ for each hydrodynamic time step.  
Because of the much larger $\Delta y$, subcycling is not necessary in 
the CGMV simulation. That adds another relative
economy to the CGMV calculation.

At the end of this simulation, $t = 20$, the modified shock
structure is approaching a dynamical equilibrium in the sense
that the postshock values of $\rho$, $P_g$ and $P_c$ will
not change much at later times. Since this shock is weaker
than the Mach 40 shocks examined earlier modifications are
more moderate. On the other hand, as
expected from the stronger momentum dependence of $x_d(p)$,
the shock precursor broadens much more quickly in the present case.
The cutoff in the CR distribution has reached roughly $p_{cutoff} \sim 10$
by $t = 20$. Longer term evolution of this shock will be
addressed below.

The agreement between the FD and CGMV solutions shown in Fig. 4 is good, 
although not as close as it was in the examples illustrated in Fig. 1 and Fig. 2. 
The more apparent distinctions between the two solutions in the present
case come from effective differences in the
application of the TL injection model with Bohm diffusion
in FD and CGMV methods.
Recall that the CGMV scheme applies the diffusion
coefficients averaged across the momentum bins 
(see equations \ref{kni}, \ref{kgi}).
The Bohm diffusion model has a very steep momentum dependence
for nonrelativistic particles; namely, $\kappa \propto p^2$.
At low momenta where injection takes place the averaging
increases the effective diffusion coefficient, and, thus, 
the leakage flux of suprathermal particles,
leading to higher injection rate compared to the FD scheme
for the same TL model parameters.
Consequently, the distribution function in the second bin at 
$p_2\approx 1.4\times10^{-2}$
is slightly higher in the CGMV scheme, as evident in Figs. 4-5.
Note that $f(p_1)$ is anchored on the tail of Maxwellian distribution. 
The CGMV solutions accordingly show slightly more efficient 
CR acceleration than the FD solutions at early times. In this
test $P_c$ is about 5 \% greater in the CGMV simulation at $t = 20$.
Since the CGMV scheme can be implemented with nonuniform momentum
bins, such differences could be reduced by making the
momentum bins smaller at low momentum in instances where the
details relating to the injection rate were important.

We show in Fig. 5 the evolution of this same shock extended to
$t = 10^3$, as computed with the CGMV method.
This simulation is computed on the domain [0,800], spanned
by a base spatial grid of $2\times 10^4$ zones, 
giving $\Delta x_0 = 4\times 10 ^{-2}$.
We also included 7 refined grid levels at the subshock, giving
$\Delta x_7 = 3.1\times 10^{-4}$.
This grid spacing is insufficient for convergence at
the injection momentum, $p_{inj} \approx 10^{-2}$, so that
the very early evolution is somewhat slower than in the simulations 
shown in Fig. 4. 
However, once shock modification becomes strong
evolution becomes roughly self-similar, 
as pointed out previously \cite{kjg02}. The time asymptotic states
do not depend sensitively on the early injection history.
The self-similar behavior results with Bohm diffusion
from a match between the
upstream and downstream extensions of the CR population.
One also sees from the form of the distribution function in Fig. 5 
that the postshock gas temperature has
stabilized, while the previously-explained concave
form to the CR distribution is better developed than it
was at earlier times.

This simulation illustrates nicely the relative efficiency of the CGMV scheme.
The equivalent FD simulation would be very much more expensive, because
this model requires a long execution time and a large spatial domain. 
With Bohm diffusion $\kappa \propto p$ for ultrarelativistic CRs,
so that the scale of the precursor, $x_d \propto p_{cutoff}$.
At the same time the peak in the CR momentum distribution
extends relatively slowly, with $p_{cutoff} \propto t$. 
The required spatial grid is, thus, 40 times longer than for the 
shorter simulation illustrated in Fig. 4. The simulated
time interval in the extended simulation was 50 times longer. 
Together those increase the total computational time by a factor 2000.
The FD calculation with $x=[0,20]$ to $t=20$ took about 2 CPU days
on our fastest available processor, so the extended simulation would
have been unrealistic using the FD method. 
The extended CGMV simulation, however, required only about 10 times the effort
of the shorter FD simulation, clearly demonstrating the efficiency of
the CGMV scheme. 
This speed-up is a result of combination of several factors:
20 times larger grid spacing, no need for subcycling for the diffusion 
convection solver, and, of course, a smaller number of momentum bins.

\section{Conclusions}

Detailed time dependent simulations of nonlinear CR shock
evolution are very expensive if one allows for inclusion of
arbitrary, self-consistent and possibly time dependent
spatial diffusion, as well as various other momentum dependent
transport processes. The principal computational cost
in such calculations is typically the CR transport
itself, and, in self-consistent calculations, the analogous 
transport of the MHD wave turbulence that mediates CR transport.
Tracking these behaviors requires adding at least one
physical dimension to the simulations compared to the associated
hydrodynamical
calculations, since the collisionless media involved are sensitive to
the phase space configurations of the particles and waves. Particle
kinetic equations (commonly the so-called diffusion
convection equation) provide a straightforward approach
to addressing this problem and can be coupled conveniently with
hydrodynamical equations that track mass and
bulk momentum and energy effectively. 

Momentum derivatives of the CR 
distribution function in the diffusion convection equation
are most frequently handled by finite differences. Although it is
simple, that approach requires moderately fine resolution in momentum
space. That is a primary reason that such calculations are costly.
Here we introduce a new scheme to solve the diffusion
convection equation based on finite volumes in momentum space
with a momentum bin spacing as much as an order of magnitude larger 
than that of the usual finite difference scheme. 
We demonstrate that this Coarse Grained Momentum finite Volume ( CGMV)
method can be used successfully to
model the evolution of strong, CR-modified shocks at
much lower computational cost than the finite difference
approach. 
The computation efficiency is greatly increased, not only because 
the number of momentum bins is smaller, but also because the required
spatial grid spacing is less demanding due to the coarse-grained averaging of the
diffusion coefficient
used in the CGMV method. In addition, larger momentum bin size
can eliminate the need of subcycling of the diffusion convection solver
that can be necessary in some instances using finite
differences in momentum. 
Thus, the combination of the CGMV scheme with AMR techniques
as developed in our CRASH code, for example, should allow more detailed
modeling of the diffusive shock acceleration
process with a strongly momentum dependent diffusion model such as
Bohm diffusion, or self-consistent treatments of CR diffusion and
wave turbulence transport.

\section*{Acknowledgments}
TWJ is supported by NSF grant AST03-07600,
by NASA grants NAG5-10774, NNG05GF57G
and by the University of Minnesota Supercomputing Institute.
HK was supported by KOSEF through the Astrophysical Research Center
for the Structure and Evolution of Cosmos (ARCSEC).



\begin{figure}
\begin{center}
\includegraphics*[height=40pc]{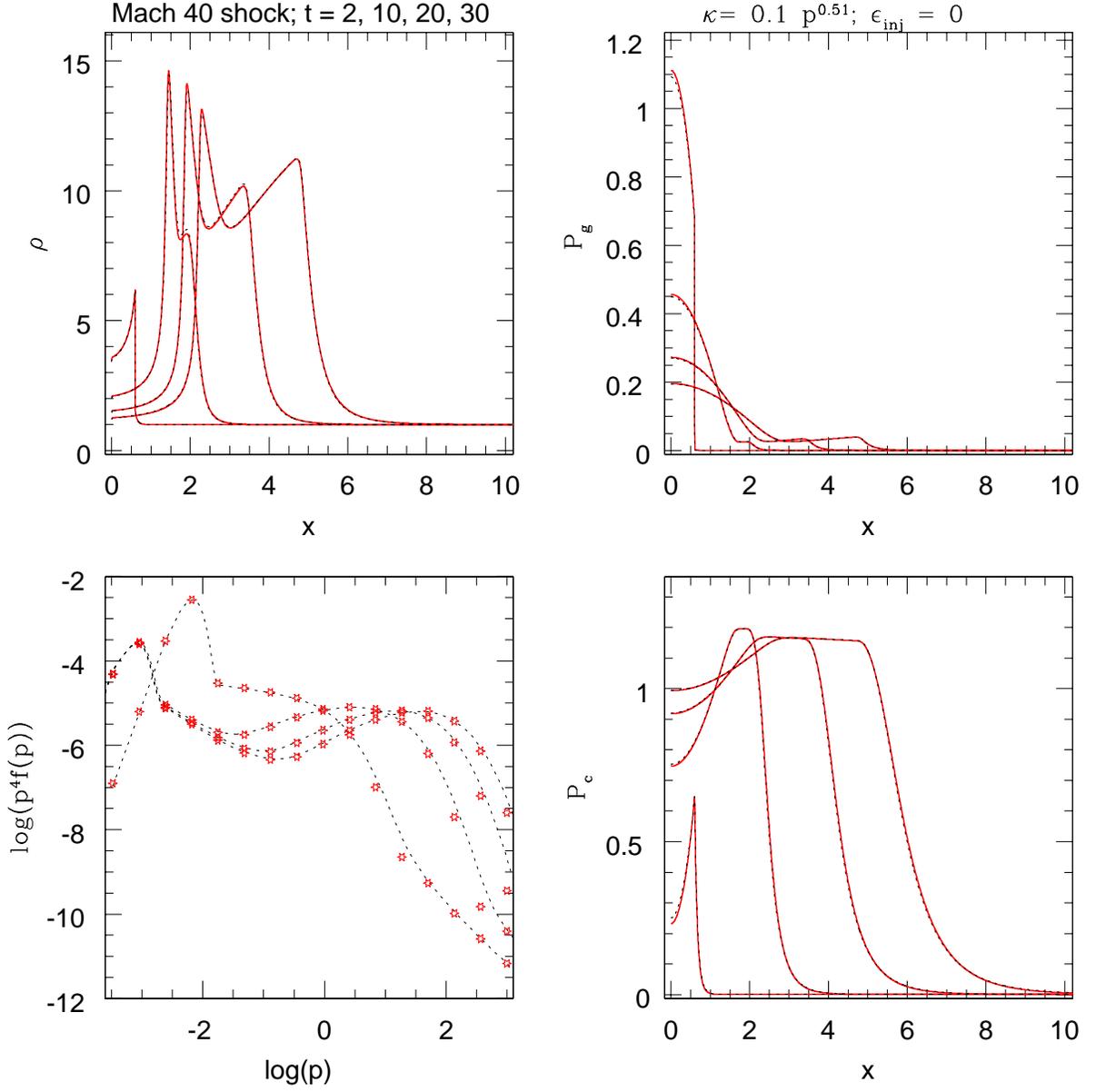}
\end{center}
\caption{Evolution of a CR modified Mach 40 plane shock using two different
schemes for solving the diffusion convection equation. The lower-left
panel illustrates the momentum distribution at the shock. The other
panels show the spatial distribution of key dynamical variables. Black
dotted lines
represent solution with a conventional finite difference scheme
using $\Delta y = 0.11$. red lines and 'stars' were obtained
using the new CGMV scheme with $\Delta y = 1.0$. 
The solutions are almost indistinguishable.
A pre-existing CR population, $f(p) \propto p^{-4.5}$, corresponding
to the upstream CR pressure, $P_c= P_g$ is included, without fresh
injection at the shock ($\epsilon_{inj}=0.$).  
}
\label{mach40a}
\end{figure}

\begin{figure}
\begin{center}
\includegraphics*[height=40pc]{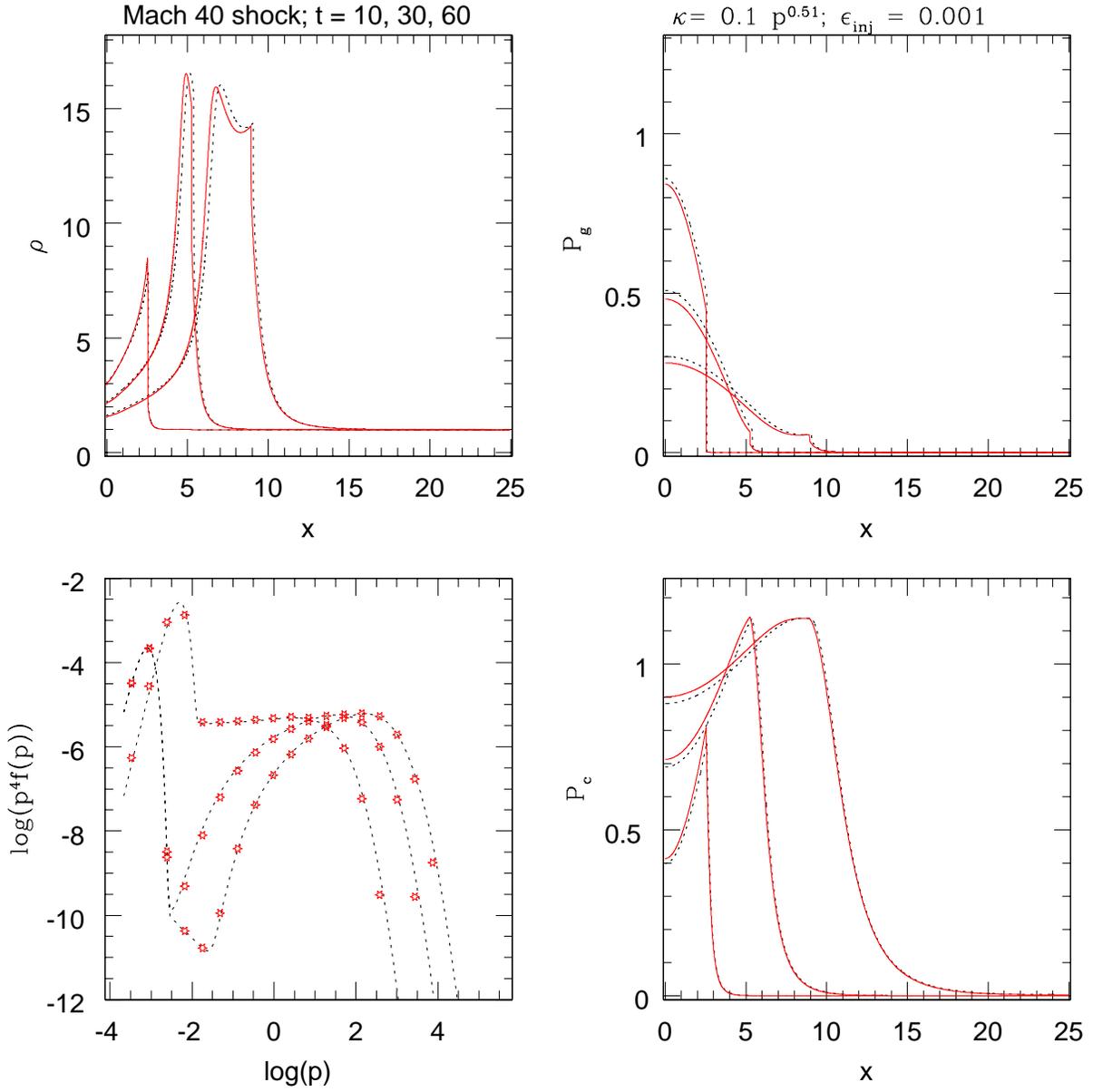}
\end{center}
\caption{Same as Fig. 1 except the flux fraction injection model is
adopted with no pre-existing CR population.
The momentum
distribution includes the Maxwellian distribution of the thermal
plasma based on its temperature.
}
\label{mach40b}
\end{figure}

\begin{figure}
\begin{center}
\includegraphics*[height=40pc]{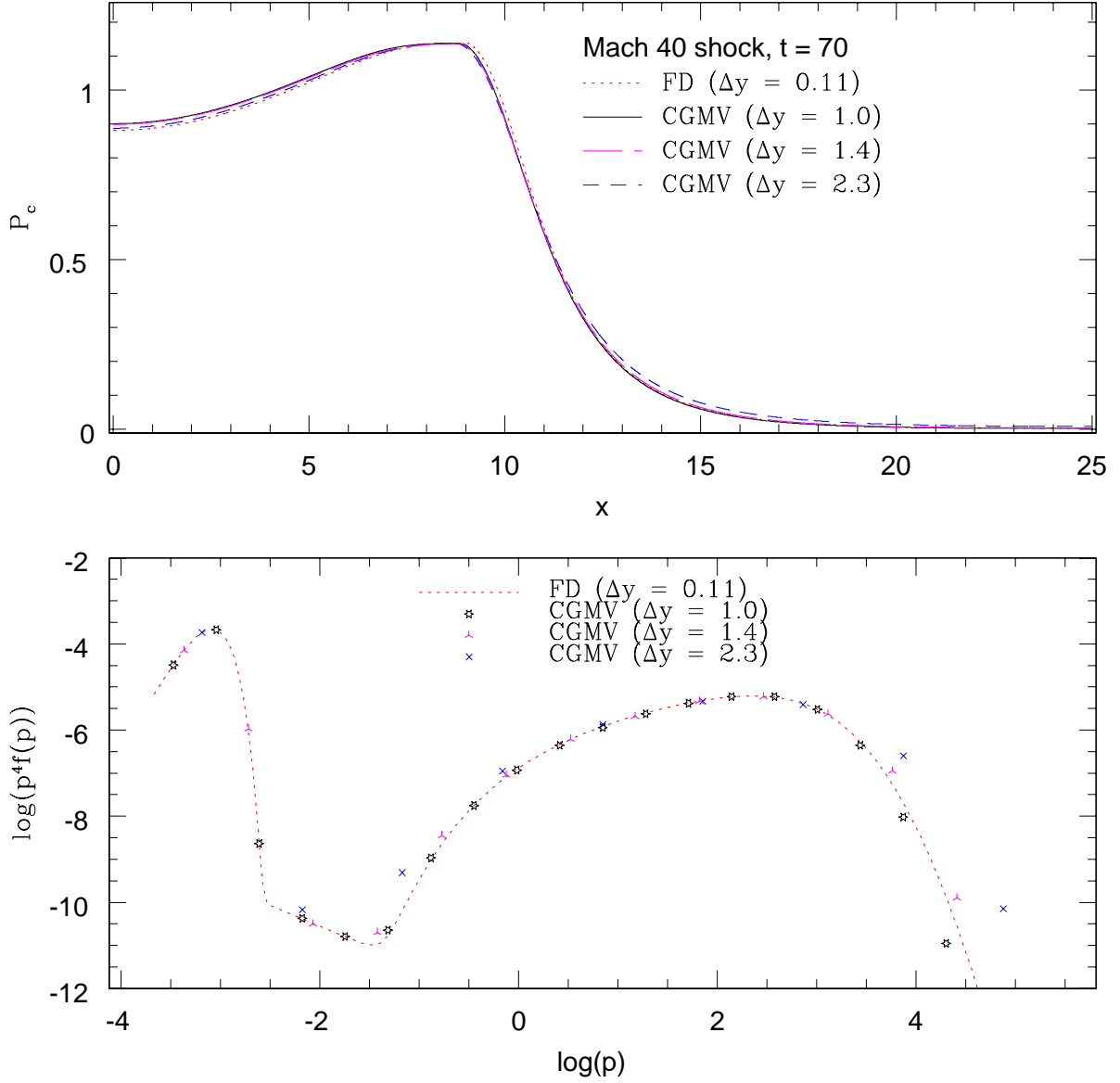}
\end{center}
\caption{Top: The spatial distribution of $P_c$ at $t = 70$ for
the same shock as shown in Fig 2. The different curves represent
results computed using the FD scheme and three different momentum
resolutions with the CGMV scheme. Bottom: The CR distribution function 
at the shock from the same simulations.}
\label{CRcomp}
\end{figure}

\begin{figure}
\begin{center}
\includegraphics*[height=40pc]{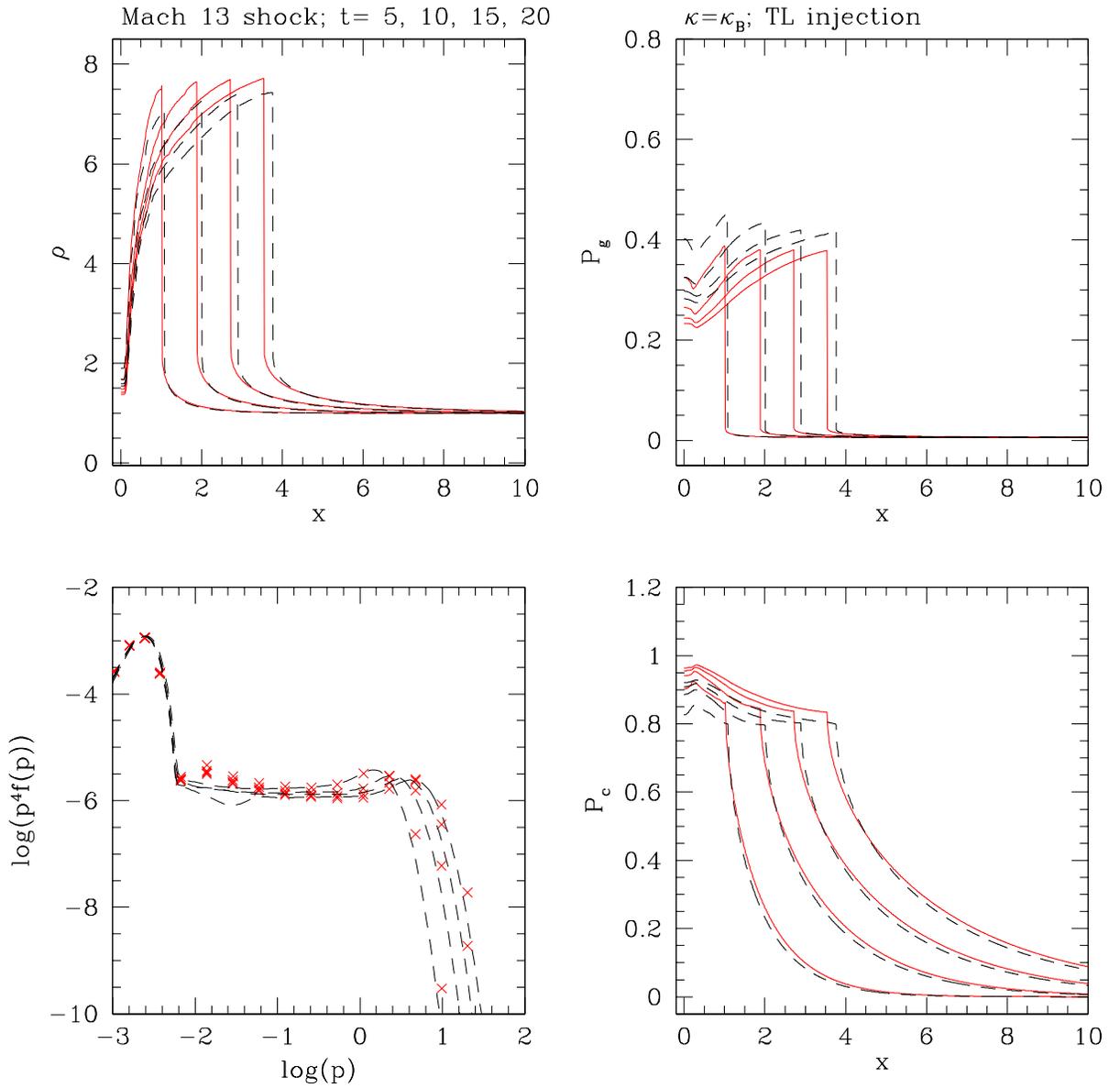}
\end{center}
\caption{Evolution of a CR modified Mach 13 plane shock using the
CRASH code with thermal leakage injection. 
The black dashed lines
represent solution with a conventional finite difference scheme
using 240 momentum points ($\Delta y = 0.058$). The red lines and 'Xs' were obtained
using the new CGMV scheme with 20 momentum bins ($\Delta y = 0.72$).}
\label{AMRmach10a}
\end{figure}

\begin{figure}
\begin{center}
\includegraphics*[height=40pc]{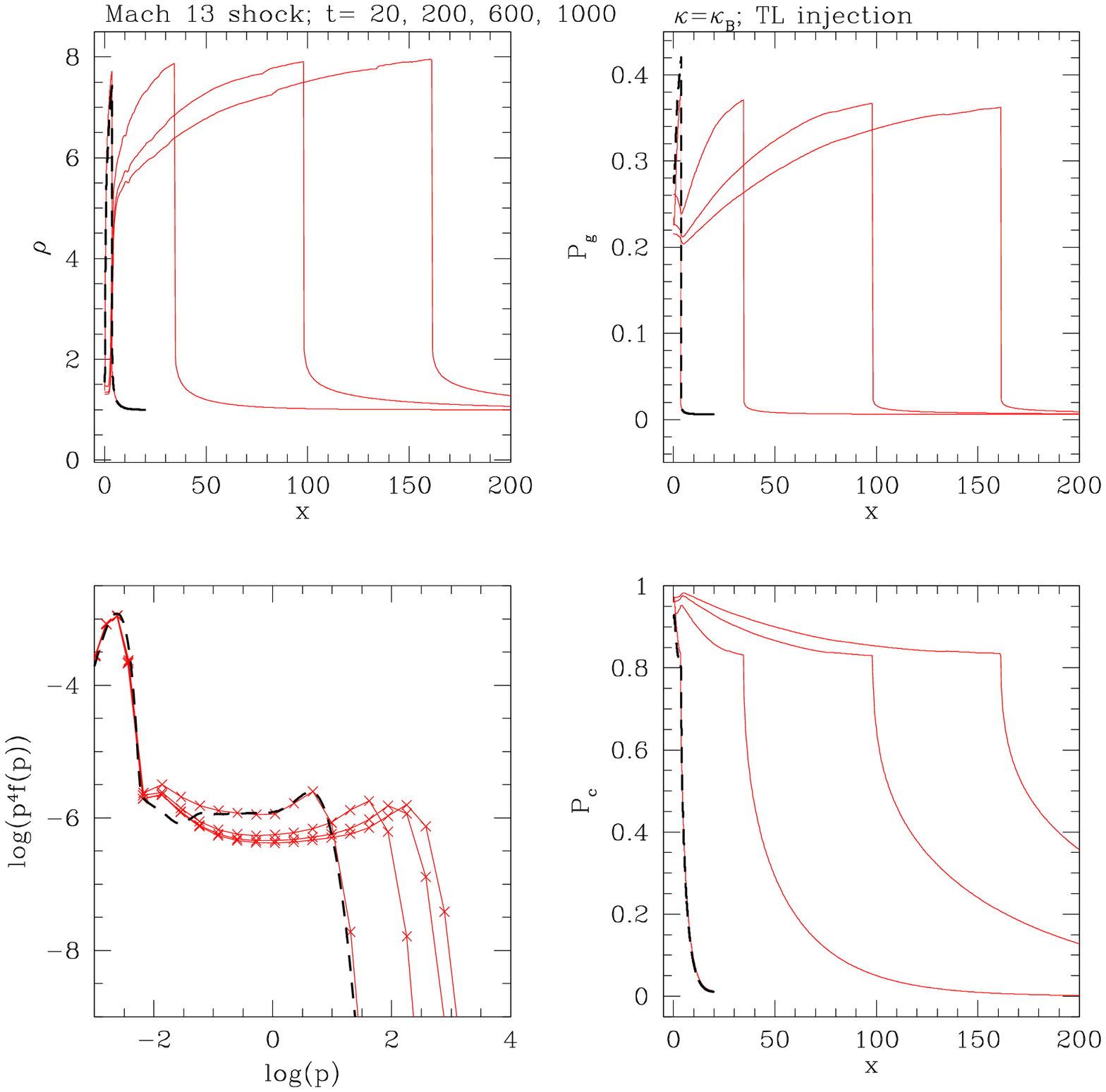}
\end{center}
\caption{Same as Fig. 4 except that the terminal time and simulation box 
are extended to $t=10^3$ and $x_{max}=800$, respectively.
The heavy dashed lines represent solution at $t=20$ with a conventional finite
difference scheme using 240 momentum points ($\Delta y = 0.058$).
The red solid lines and 'X's represent CGMV solutions at $t=20, 200, 600,$
and 1000 with 20 momentum bins ($\Delta y = 0.72$).}
\label{AMRmach10b}
\end{figure}


\begin{thebibliography}{00}




\bibitem{ach82}
Achterberg A. Astron. \& Astrophys. 98 (1982) 195

\bibitem{bell78}
Bell, A. R. M.N.R.A.S. 182 (1978) 147

\bibitem{berz94} 
Berezhko E.G., Yelshin V.K., \& Ksenofontov L.T. Astroparticle Physics 
2 (1994) 215

\bibitem{berz95}
Berezhko E.G., Ksenofontov L.T., \& Yelshin V.K. Nuclear Physics B 39a (1995) 171

\bibitem{blas02}
Blasi, P., Astroparticle Physics 16 (2002) 429

\bibitem{dv81} 
Drury, L.~O'C., \& V\"olk, H.~J., Astrophys. J. 248 (1981) 344

\bibitem{dru83} 
Drury, L.~O'C. Rept. Prog. Phys. 46 (1983) 973

\bibitem{drufal86} 
Drury, L.~O'C., \& Falle, S.~A.~E.~G. M.N.R.A.S. 223 (1986) 353

\bibitem{ell84}
Ellison, D.~C. \& Eichler, D. Astrophys. J. (1984) 691

\bibitem{EMP90}
Ellison, D.~C., M\"obius, E., \& Paschmann, G. Astrophys. J. 352 (1990) 376

\bibitem{EBS93}
Ellison, D.~C., Giacalone, J, Burgress D., \& Schwartz, S.~J. 
JGR 98 (1993) 21085 

\bibitem{fg87}
Falle, S. A. E. \& Giddings, J. R. M.N.R.A.S. 225 (1987) 399

\bibitem{fjr95}
Frank, A., Jones, T. W. \& Ryu, D., Astrophys. J. 441 (1995) 629

\bibitem{giaetal93}
Giacalone, J., Burgess, D., Schwartz, S. J. \& Ellison, D. C.
Astrophs. J. 402 (1993) 550

\bibitem{gies00}
Gieseler, U. D. J., Jones, T. W. \& Kang, H. Astron. \& Astrophys. 364 (2000) 911

\bibitem{hart83}
Harten, A. J. Comp. Phys. 49 (1983) 357
\bibitem{jon93}
Jones, T.~W. Astrophys. J. 619 (1993) 619

\bibitem{jon94}
Jones, T.~W. Astrophys. J. Supplements 90 (1994) 969

\bibitem{jre99}
Jones, T. W., Ryu, D. \& Engel, A., Astrophys. J. 512 (1999) 105

\bibitem{jj99}
Jun, B.-I. \& Jones, T. W., Astrophys. J. 511 (1999) 774

\bibitem{kj91}
Kang, H. \& Jones, T. W., M.N.R.A.S. 249 (1991) 439

\bibitem{kj95}
Kang, H. \& Jones, T. W., Astrophys. J. 447 (1995) 944

\bibitem{kj97}
Kang, H. \& Jones, T. W., Astrophys. J. 476 (1997) 875

\bibitem{kjls01} 
Kang, H., Jones, T. W., LeVeque, R. J., and Shyue, K. M. Astrophys. J. 550 
(2001) 737

\bibitem{kjg02} 
Kang, H., Jones, T. W., \& Gieseler, U.D.J. Astrophys. J. 579 (2002) 337

\bibitem{lag83}
Lagage, P. O. \& Cesarsky, C. J. Astron. \& Astrophys. 125 (1983) 249

\bibitem{levshy95} 
LeVeque, R. J., and Shyue, K. M. SIAM J. Scien. Comput. 16 (1995) 348

\bibitem{mal97}
Malkov, M.A. Astrophys. J. 485 (1997) 638

\bibitem{mal99}
Malkov, M.A. Astrophys. J. 511 (1999) L53

\bibitem{malvol98} 
Malkov M.A., and V\"olk H.J. Adv. Space Res. 21 (1998) 551

\bibitem{maldru01} 
Malkov M.A., \& Drury, L.O'C. Rep. Progr. Phys. 64 (2001) 429

\bibitem{marc99}
Marcowith, A. \& Kirk, J. G. Astron. \& Astrophys. 347 (1999) 391

\bibitem{min01}
Miniati, F. Comp. Phys. Comm. 141 (2001) 17

\bibitem{skill75}
Skilling, J., M.N.R.A.S., 223 (1975) 353

\bibitem{scho90} 
Scholar, M. Geophys Res Lett. 17 (1990) 11

\end{thebibliography}
\end{document}